%% file: paper.tex
\documentclass[11pt]{article}
\usepackage[margin=1in]{geometry}
\usepackage{amsmath,amssymb}
\usepackage{graphicx}
\usepackage{booktabs}
\usepackage{hyperref}
\usepackage[numbers]{natbib}
\usepackage{placeins}
\usepackage{float}
\usepackage{tikz}
\usetikzlibrary{arrows.meta,positioning,fit}

\title{PredictionMarketBench: A SWE-bench-Style Framework for Backtesting Trading Agents on Prediction Markets}
\author{%
  Avi Arora\\\texttt{avi@oddpool.com}\thanks{Oddpool: \href{https://oddpool.com}{oddpool.com} \;\textbar\; Code: \href{https://github.com/Oddpool/PredictionMarketBench}{PredictionMarketBench}.}
  \and
  Ritesh Malpani\\\texttt{ritesh.malpani7@gmail.com}
}
\date{}

\begin{document}
\maketitle

\begin{abstract}
Prediction markets offer a natural testbed for trading agents: contracts have binary payoffs, prices can be interpreted as probabilities, and realized performance depends critically on market microstructure, fees, and settlement risk. We introduce \textbf{PredictionMarketBench}, a SWE-bench-style benchmark for evaluating algorithmic and LLM-based trading agents on prediction markets via deterministic, event-driven replay of historical limit-order-book and trade data.

PredictionMarketBench standardizes (i) episode construction from raw exchange streams (orderbooks, trades, lifecycle, settlement), (ii) an execution-realistic simulator with maker/taker semantics and fee modeling, and (iii) a tool-based agent interface that supports both classical strategies and tool-calling LLM agents with reproducible trajectories. We release four Kalshi-based episodes spanning cryptocurrency, weather, and sports.

We report baseline results for a RandomAgent, a tool-calling LLM agent (\texttt{gpt-4.1-nano}), and a classic Bollinger Bands mean-reversion strategy, illustrating that naive activity can underperform due to transaction costs and settlement losses, while fee-aware algorithmic alphas can remain competitive in volatile episodes.
\end{abstract}

\section{Introduction}
\input{sections/01_introduction}

\section{PredictionMarketBench}
\input{sections/02_predictionmarketbench}

\section{Data}
\input{sections/03_data}

\section{Agents}
\input{sections/04_agents}

\section{Experiments}
\input{sections/05_experiments}

\section{Results}
\input{sections/06_results}

\section{Related Work}
\input{sections/07_related_work}

\section{Discussion and Conclusion}
\input{sections/08_discussion}

\appendix
\section{LLM system prompt}
\label{app:system-prompt}
\input{sections/A_system_prompt}

\bibliographystyle{plainnat}
\bibliography{refs}

\end{document}

%% file: sections/01_introduction.tex
Prediction markets are exchange-traded contracts whose prices can often be interpreted as collective probabilistic forecasts about real-world events. Under appropriate design and participation, they can aggregate dispersed information and yield accurate predictions \citep{wolfers2004prediction,arrow2008promise}. Empirically, election prediction markets such as the Iowa Electronic Markets have been shown to outperform traditional polls at longer horizons \citep{berg2008accuracy}, while the broader literature highlights the importance of market design and susceptibility to manipulation in determining forecast quality \citep{bergriet2014design}.

A key practical feature of modern prediction markets is that liquidity may be thin or fragmented across many related questions. Automated market makers and cost-function market makers (e.g., LMSR) provide continuous pricing under sparse order flow and have become foundational to the study of prediction market design \citep{hanson2003lmsr}. At the same time, many venues (including regulated exchanges) operate as electronic limit-order markets, where execution outcomes depend on microstructure: spreads, queue position, and fees \citep{ohara1995microstructure}.

These properties make prediction markets a compelling but challenging environment for algorithmic and AI trading agents. Agents must reason under discrete settlement payoffs, horizon-dependent risk, and high transaction costs; furthermore, naive backtests can be misleading due to selection effects and overfitting \citep{bailey2017backtest}. Prior work has studied prediction markets through agent-based modeling and trader behavior, emphasizing how heterogeneous beliefs and bounded-rational strategies map into aggregate price dynamics \citep{yuchen2011abm}.

Recently, LLM-based and agentic AI systems have begun to be applied to prediction markets and related financial tasks, for example by using natural-language understanding to cluster market questions and derive tradable signals \citep{capponi2025semantic}. However, systematic evaluation remains difficult: results often depend on dataset choice, execution assumptions, and opaque experimental details.

This paper introduces \textbf{PredictionMarketBench}, an event-driven benchmark framework for backtesting trading agents on prediction-market data with execution-realistic replay. The design is inspired by harness-first evaluation approaches in other agent domains---most notably SWE-bench, which evaluates agents against standardized task instances with an executable harness \citep{jimenez2024swebench}. PredictionMarketBench adapts this idea to trading: we define standardized \emph{episodes} (events), a deterministic replay \emph{simulator} over historical market data streams, and a consistent set of metrics for apples-to-apples comparisons across heterogeneous agent implementations.

Our primary contributions are:
\begin{itemize}
  \item An episode construction pipeline that converts raw exchange data into self-contained benchmark instances (metadata, orderbooks, trades, settlements).
  \item A deterministic, event-driven simulator supporting maker/taker execution semantics and fee modeling appropriate for binary contracts.
  \item An agent interface and benchmark harness that standardize evaluation, logging, and metric computation for both algorithmic and tool-calling LLM agents.
\end{itemize}

%% file: sections/02_predictionmarketbench.tex
PredictionMarketBench is a replay-based benchmark for evaluating trading agents on \emph{binary event contracts} (YES/NO) listed on prediction markets. Each benchmark instance is an \emph{episode} corresponding to a single underlying event (e.g., a price threshold, a weather outcome, or a sports result). The benchmark couples (i) an episode format that makes historical market data portable and self-contained and (ii) a deterministic, event-driven simulator that replays historical market microstructure.

The current benchmark targets data collected from Kalshi, a U.S. prediction market regulated by the Commodity Futures Trading Commission (CFTC) as a Designated Contract Market (DCM) \citep{kalshi_regulation,cftc_kalshi_dcm}.

\subsection{Benchmark Construction}
\label{sec:benchmark-construction}

\paragraph{Data collection.}
The benchmark is built from three primary historical streams:
\begin{itemize}
  \item \textbf{Orderbook updates} capturing limit-order-book depth over time.
  \item \textbf{Trade prints} capturing executed transactions (used to model maker fills).
  \item \textbf{Market lifecycle and settlement} events capturing trading halts/closures and final outcomes.
\end{itemize}
These streams are aligned to a unified UTC timeline and then grouped by event identifier into episodes.

\paragraph{Episode representation.}
Each episode is stored as a directory containing:
\begin{itemize}
  \item \texttt{metadata.json} (episode configuration, tickers, time bounds, bankroll, execution mode, fee model version),
  \item \texttt{orderbook.parquet} (time-series orderbook snapshots),
  \item \texttt{trades.parquet} (historical trades; when available),
  \item \texttt{settlement.json} (final YES/NO outcome per ticker).
\end{itemize}
This format is designed to be (i) \emph{portable} (each episode is self-contained), (ii) \emph{replayable} (the simulator depends only on the episode directory), and (iii) \emph{extensible} (new fields/streams can be added without invalidating older episodes).

\paragraph{Harness and simulator.}
The benchmark harness loads episodes, initializes the simulator with a fixed configuration, runs an agent through each episode under a per-step tool-call budget, and collects standardized outputs (trade logs, equity curves, and per-episode metrics). The simulator processes market events in timestamp order (using sequence numbers to break ties), updates the orderbook state, applies the execution model, and resolves contracts at settlement.

\begin{figure}[t]
\centering
\begin{tikzpicture}[
  font=\small,
  box/.style={draw, rounded corners, align=center, inner sep=9pt, text width=6.2cm},
  smallbox/.style={draw, rounded corners, align=center, inner sep=9pt, text width=6.2cm},
  arrow/.style={-{Latex[length=3mm]}, thick}
]

% Main stack (vertical)
\node[box] (episodes) {Episode files\\\texttt{metadata} / \texttt{orderbook} / \texttt{trades} / \texttt{settlement}};
\node[box, below=12mm of episodes] (harness) {BenchmarkHarness\\(iterates episodes)};
\node[box, below=12mm of harness, yshift=-6mm] (sim) {Event-driven simulator\\(replay + execution)};
\node[box, below=12mm of sim] (portfolio) {Portfolio\\cash + positions\\fees + P\&L};

% Agent side (more breathing room, slightly lower to avoid label collisions)
\node[box, right=12mm of sim, yshift=2mm, text width=4.2cm] (agent) {Agent\\policy / LLM / rules};
\node[smallbox, below=12mm of agent, text width=4.2cm] (ctx) {AgentContext (tools)\\orderbook, positions,\\place/cancel};

% Flow arrows (vertical)
\draw[arrow] (episodes) -- node[right, pos=0.55, xshift=2mm]{load} (harness);
\draw[arrow] (harness) -- node[left, pos=0.55, xshift=-2mm, yshift=3mm, fill=white, inner sep=1pt]{init + run} (sim);
\draw[arrow] (sim) -- node[right, pos=0.55, xshift=2mm]{fills, MTM} (portfolio);

% Agent interaction loop (split attachment points on simulator to avoid arrow overlap)
\draw[arrow] ([yshift=4mm]sim.east) -- node[above, yshift=1mm, fill=white, inner sep=1pt]{invoke} ([yshift=4mm]agent.west);
\draw[arrow] (agent) -- node[right, xshift=1mm]{actions (orders)} (ctx);
\draw[arrow] ([yshift=-4mm]ctx.west) to[out=180,in=0] node[below, pos=0.55, yshift=-1mm, fill=white, inner sep=1pt]{observations / API} ([yshift=-4mm]sim.east);

% Reporting back to harness
\draw[arrow] (portfolio.west) |- ++(-1.5,0) |- node[left, pos=0.2, xshift=-2mm, fill=white, inner sep=1pt]{metrics + logs} (harness.west);

% Execution loop annotation (place label just above the dashed border)
\node[draw, dashed, rounded corners, fit=(sim)(agent)(ctx), inner sep=11pt] (loop) {};
\node[font=\small, anchor=south, fill=white, inner sep=1pt] at ([yshift=1mm]loop.north) {Execution loop (cadence $\Delta t$)};

\end{tikzpicture}
\caption{PredictionMarketBench execution flow. The harness iterates over episodes, the simulator replays historical events, and the agent interacts through a Kalshi-like context API at a fixed cadence.}
\label{fig:pmbench-loop}
\end{figure}
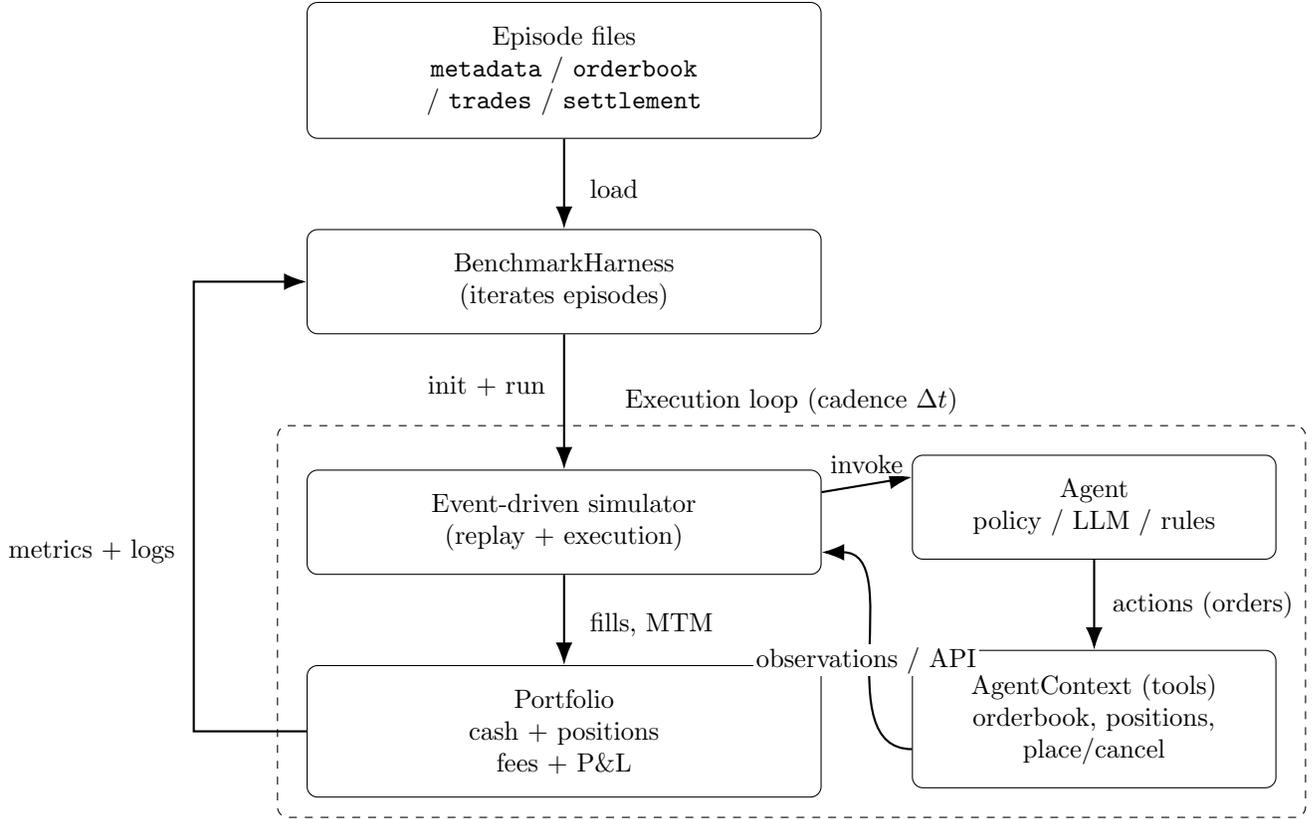

\subsection{Task formulation}
\label{sec:task-formulation}

We formulate each episode as a finite-horizon sequential decision problem over a historical timeline.

\paragraph{Decision points and termination.}
The agent is invoked at a fixed cadence (e.g., every 5 seconds). Between invocations, the simulator advances by processing all market events up to the next decision time. The episode terminates once settlement is processed and all positions are marked-to-settlement.

\paragraph{Observations.}
At each decision point the agent may query: (i) market summaries (best bid/ask per ticker), (ii) full orderbooks (optionally depth-limited), (iii) positions and cash/equity, and (iv) open resting orders.

\paragraph{Actions.}
Agents act by submitting orders defined by side (YES/NO), direction (buy/sell), order type (market/limit), size, and time-in-force (IOC/GTC/post-only, depending on execution mode).

\subsection{Features of PredictionMarketBench}
\label{sec:features}

\paragraph{Binary contract semantics.}
Binary contracts pay \$1 if the event occurs and \$0 otherwise (and analogously for NO). This structure supports a probability interpretation of prices and makes prediction markets a natural domain for studying belief formation and information aggregation \citep{wolfers2004prediction}.

\paragraph{Maker/taker execution realism.}
PredictionMarketBench supports both a taker-only mode (orders must cross immediately) and a maker-taker mode in which resting limit orders join a queue behind existing displayed volume. Maker/taker fee schedules are widely used by electronic venues to incentivize liquidity provision \citep{maker_taker}.

\paragraph{Fee-aware evaluation.}
Fees are applied per fill and accounted for in P\&L and equity curves. Explicit fee modeling is essential because transaction costs can dominate marginal strategy improvements, especially in binary contracts where notional amounts are bounded.

\paragraph{Deterministic, event-driven replay.}
All state transitions are driven by recorded market events (orderbook updates, trades, settlement). Given the same episode files, simulator configuration, and agent code (with fixed seeds), runs are deterministic, enabling reproducible comparisons and regression testing.

\paragraph{Standardized outputs and metrics.}
The harness produces trade logs, equity curves, and per-episode/aggregate metrics including P\&L, drawdown, Sharpe ratio, fees, slippage, and fill ratio.

%% file: sections/03_data.tex
The initial public release of PredictionMarketBench contains \textbf{4 episodes} collected from Kalshi in January 2026, covering three event types: \textbf{cryptocurrency} (Bitcoin daily high), \textbf{weather} (NYC high temperature), and \textbf{sports} (college football and NFL outcomes). Table~\ref{tab:episodes} summarizes each episode and its size.

\vspace{0.5em}
\begin{table}[!htbp]
\centering
\small
\begin{tabular}{@{}llrrrr@{}}
\toprule
Episode ID & Domain & Tickers & Duration (hrs) & Orderbook snaps & Trades \\
\midrule
KXBTCD-26JAN2017 & Crypto & 23 & 37.4 & 311{,}998 & 6{,}283 \\
KXHIGHNY-26JAN20 & Weather & 6 & 37.4 & 50{,}231 & 8{,}044 \\
KXNCAAF-26 & Sports & 2 & 37.4 & 8{,}320 & 171{,}786 \\
KXNFLGAME-26JAN11BUFJAC & Sports & 2 & 67.4 & 8{,}047 & 111{,}160 \\
\midrule
\textbf{Total} & --- & \textbf{33} & --- & \textbf{378{,}596} & \textbf{297{,}273} \\
\bottomrule
\end{tabular}
\caption{Episodes included in the PredictionMarketBench initial release.}
\label{tab:episodes}
\end{table}

%% file: sections/04_agents.tex
PredictionMarketBench exposes a lightweight \texttt{Agent} interface designed to support both traditional algorithmic strategies (e.g., rule-based, market-making, statistical arbitrage) and LLM-driven agents that reason over the current state and call tools.

Agents interact with the environment exclusively through an \texttt{AgentContext} that provides Kalshi-like API primitives (e.g., querying markets and orderbooks, inspecting positions and cash, and placing/canceling orders). This design decouples strategy logic from simulator internals and enables plug-and-play evaluation of heterogeneous agent implementations under a shared harness.

All observations and actions are timestamped on a unified UTC timeline, allowing agents to align market state with external data sources during replay. For example, in the weather episode(s), an agent can query a historical weather API at the episode timestamp to produce exogenous features, while preserving strict temporal consistency with the market data.

%% file: sections/05_experiments.tex
\subsection{LLM agent}
\label{sec:exp-llm}

We evaluate a tool-calling LLM trading agent that operates in a stateless per-step loop: at each decision time, the harness constructs a textual summary of the current state (cash, equity, time-to-settlement, top-of-book quotes, and current positions), and the model responds with zero or more tool calls.

The agent is provider-agnostic and interacts only through the benchmark \texttt{AgentContext} interface (market data queries, portfolio queries, and order placement/cancellation). The full production system prompt is provided in Appendix~\ref{app:system-prompt}. Unless otherwise stated, we use a deterministic decoding configuration (temperature 0.0).

\paragraph{Run configuration.}
\begin{table}[!htbp]
\centering
\small
\begin{tabular}{@{}ll@{}}
\toprule
Parameter & Value \\
\midrule
Model & \texttt{gpt-4.1-nano} \\
Temperature & 0.0 \\
Max tool rounds per step & 3 \\
Agent call cadence & 300s (5 minutes) \\
Equity sampling interval & 60s \\
Maker queue mode & \texttt{trade\_only} \\
\bottomrule
\end{tabular}
\caption{LLM agent production-run configuration.}
\label{tab:llm-config}
\end{table}

\paragraph{Logging and reproducibility.}
We log full trajectories (state messages, tool calls, and tool results) together with per-step timestamps to enable exact replay and post-hoc analysis. The production runner checkpoints the trajectory every 50 steps and records aggregate outputs (trade logs and equity curves).

\subsection{Random agent baseline}
\label{sec:exp-random}

As a low-complexity baseline, we include a RandomAgent that (with probability 0.1 per step) selects a random quoted market and submits a small market order (1--3 contracts) on a random side, subject to a per-ticker position cap of \(\pm 10\) contracts. The RNG is seeded per-episode (via the episode identifier) for determinism. The random baseline uses the same simulator cadence (300s) and observation interface as the LLM agent.

\subsection{Bollinger Bands (mean-reversion alpha)}
\label{sec:exp-bollinger}

We also evaluate a classic Bollinger Bands mean-reversion strategy \citep{bollinger2001} adapted to binary contract prices. At each step, for each ticker, the agent maintains a rolling mid-price history, computes a simple moving average (period 20) and standard-deviation bands (\(k=2\)), and trades only on band crossings (buy near the lower band; sell/hedge near the upper band). To reduce fee impact, the strategy primarily uses maker \texttt{GTC} limit orders.

%% file: sections/06_results.tex
We report results for three baselines: a tool-calling LLM agent (\texttt{gpt-4.1-nano}), a simple RandomAgent, and a classic Bollinger Bands mean-reversion strategy \citep{bollinger2001}. All are evaluated on the same four episodes with \$1{,}000 initial bankroll per episode and a 5-minute agent cadence.

\subsection{Per-episode results}

\begin{table}[H]
\centering
\small
\begin{tabular}{@{}lrrrrr@{}}
\toprule
Episode & PnL (\$) & Return (\%) & Max DD (\%) & Contracts & Fill (\%) \\
\midrule
\multicolumn{6}{l}{\textbf{RandomAgent}}\\
KXBTCD-26JAN2017 & -2.89 & -0.29 & 0.83 & 32 & 100.0 \\
KXHIGHNY-26JAN20 & -0.75 & -0.07 & 0.78 & 33 & 100.0 \\
KXNCAAF-26 & +0.62 & +0.06 & 0.08 & 3 & 100.0 \\
KXNFLGAME-26JAN11BUFJAC & -2.11 & -0.21 & 0.44 & 15 & 100.0 \\
\midrule
\textbf{Total} & \textbf{-5.13} & \textbf{-0.13} & --- & \textbf{83} & \textbf{100.0} \\
\bottomrule
\end{tabular}
\caption{RandomAgent performance by episode (run date: 2026-01-27).}
\label{tab:random-per-episode}
\end{table}
\vspace{-0.75em}

\begin{table}[H]
\centering
\small
\begin{tabular}{@{}lrrrrr@{}}
\toprule
Episode & PnL (\$) & Return (\%) & Max DD (\%) & Contracts & Fill (\%) \\
\midrule
\multicolumn{6}{l}{\textbf{LLM agent (gpt-4.1-nano)}}\\
KXBTCD-26JAN2017 & -64.79 & -6.48 & 36.0 & 1{,}318 & 100.0 \\
KXHIGHNY-26JAN20 & +7.49 & +0.75 & 3.4 & 234 & 71.6 \\
KXNCAAF-26 & -34.07 & -3.41 & 3.6 & 145 & 100.0 \\
KXNFLGAME-26JAN11BUFJAC & -19.25 & -1.93 & 3.0 & 380 & 66.2 \\
\midrule
\textbf{Total} & \textbf{-110.62} & \textbf{-2.77} & --- & \textbf{2{,}077} & \textbf{84.4} \\
\bottomrule
\end{tabular}
\caption{LLM agent performance by episode (run dates: 2026-01-25 to 2026-01-26).}
\label{tab:llm-per-episode}
\end{table}
\vspace{-0.75em}

\begin{table}[H]
\centering
\small
\begin{tabular}{@{}lrrrrr@{}}
\toprule
Episode & PnL (\$) & Return (\%) & Max DD (\%) & Contracts & Fill (\%) \\
\midrule
\multicolumn{6}{l}{\textbf{Bollinger Bands}}\\
KXBTCD-26JAN2017 & 64.10 & 6.41 & 3.18 & 616 & 94.5 \\
KXHIGHNY-26JAN20 & 0.87 & 0.09 & 0.30 & 80 & 100.0 \\
KXNCAAF-26 & 0.00 & 0.00 & 0.00 & 0 & 100.0 \\
KXNFLGAME-26JAN11BUFJAC & 1.71 & 0.17 & 0.05 & 18 & 100.0 \\
\midrule
\textbf{Total} & \textbf{66.68} & \textbf{1.67} & 3.18 & \textbf{714} & \textbf{98.6} \\
\bottomrule
\end{tabular}
\caption{Bollinger Bands performance by episode (mean-reversion alpha).}
\label{tab:bollinger-per-episode}
\end{table}

\FloatBarrier
\subsection{Aggregate comparison}

\begin{table}[!htbp]
\centering
\small
\begin{tabular}{@{}lrrrrrr@{}}
\toprule
Agent & Total PnL (\$) & Return (\%) & Max DD (\%) & Contracts & Orders & Fill (\%) \\
\midrule
Bollinger Bands & 66.68 & 1.67 & 3.18 & 714 & --- & 98.6 \\
LLM (gpt-4.1-nano) & -110.62 & -2.77 & 36.0 & 2{,}077 & 603 & 84.4 \\
RandomAgent & -5.13 & -0.13 & 0.83 & 83 & 45 & 100.0 \\
\bottomrule
\end{tabular}
\caption{Aggregate comparison across all four episodes.}
\label{tab:results-compare}
\end{table}

\FloatBarrier

Bollinger Bands achieves positive overall P\&L, with most profit concentrated in the volatile Bitcoin threshold episode. In contrast, the LLM agent trades far more aggressively and suffers large settlement losses; the RandomAgent loses less primarily due to low trading intensity.

%% file: sections/07_related_work.tex
\paragraph{Prediction markets and market design.}
Prediction markets have been studied as mechanisms for aggregating dispersed information into probabilistic forecasts \citep{wolfers2004prediction,arrow2008promise}. Empirical work on the Iowa Electronic Markets documents long-horizon accuracy and emphasizes how design choices and manipulation incentives affect information quality \citep{berg2008accuracy,bergriet2014design}. On the market-design side, automated market makers and cost-function approaches such as LMSR provide continuous pricing under sparse order flow \citep{hanson2003lmsr}.

\paragraph{Market microstructure and execution.}
Because many venues operate as limit-order markets, agent performance depends on microstructure effects such as bid--ask spreads, queue priority, and fees \citep{ohara1995microstructure}. These considerations motivate evaluation harnesses that explicitly model execution and transaction costs rather than relying on idealized mid-price fills.

\paragraph{Learning and trading agents in market environments.}
A broad line of work studies learning agents in market settings, including reinforcement learning for trading and price discovery in limit-order markets \citep{he2019rl_lom} and agent-based modeling of prediction markets \citep{yuchen2011abm}. In financial ML more generally, open-source RL pipelines such as FinRL aim to standardize environments, algorithms, and evaluation for trading tasks \citep{liu2021finrl}.

\paragraph{Simulators and benchmarks for agent evaluation.}
High-fidelity market simulators such as ABIDES were developed explicitly to support AI-agent research in market applications \citep{byrd2019abides}. Building on such simulators, recent work trains RL agents for optimal execution in realistic limit-order-book environments \citep{karpe2020marl}. PredictionMarketBench is complementary: rather than learning endogenous market dynamics, it focuses on deterministic replay of real prediction-market microstructure (orderbooks, trades, settlement) with maker/taker semantics and fee modeling to enable apples-to-apples backtesting.

\paragraph{LLM-based and agentic systems.}
Harness-first evaluation has become standard in other agent domains (e.g., SWE-bench) to ensure reproducible, instance-based comparisons \citep{jimenez2024swebench}. In prediction markets, recent work explores agentic AI and semantic structure discovery as a basis for trading signals \citep{capponi2025semantic}. Our benchmark provides a concrete experimental substrate for such agents by standardizing the observation/action interface, tool budgets, and deterministic replay.

%% file: sections/08_discussion.tex
\subsection{Limitations}
\label{sec:limitations}

PredictionMarketBench is an initial step toward standardized, execution-aware evaluation for prediction-market trading agents, and it has several limitations. First, the current release is small (4 episodes from January 2026) and spans a limited set of event types; broader statistical claims require more diverse markets and time periods. Second, while maker/taker execution and fees are modeled, the simulator is still an abstraction of live trading: latency, exchange-specific priority rules, and strategic interaction with other agents are not modeled explicitly \citep{ohara1995microstructure}. Third, our baseline agents are intentionally simple and are not tuned for out-of-sample robustness; repeated iteration on this dataset risks backtest overfitting \citep{bailey2017backtest}.

\subsection{Conclusion}
\label{sec:conclusion}

We presented PredictionMarketBench, a SWE-bench-style benchmark for backtesting algorithmic and LLM-based trading agents on prediction markets using deterministic replay of real orderbooks, trades, and settlement outcomes. Our initial results illustrate the importance of execution realism and transaction costs: an active LLM agent can incur large settlement losses, while a fee-aware algorithmic alpha (Bollinger Bands) can remain profitable in volatile episodes. We release the benchmark to support reproducible comparisons and encourage future work on expanding the dataset, adding stronger baselines, and studying robust agent design under realistic execution constraints.

%% file: sections/A_system_prompt.tex
\begin{verbatim}
You are a trading agent operating on Kalshi prediction markets.

## Objective
Maximize profit by trading binary event contracts. Each contract pays $1.00 (100 cents) if the event outcome matches (YES wins = $1 for YES holders, NO wins = $1 for NO holders).

## Market Mechanics
- Prices are quoted in cents (1-99)
- A YES contract at 60c means the market implies ~60% probability
- A NO contract at the same market would be at ~40c (100 - 60 = 40)
- You can BUY or SELL on either YES or NO side

## Fee Structure (Critical for Profitability)
- **Taker fee**: 7% × price × (1 - price/100) - applies to market orders and limit orders that cross the spread
- **Maker fee**: 1.75% × price × (1 - price/100) - applies to resting limit orders that provide liquidity
- Example: At 50c, taker fee = 1.75c per contract, maker fee = 0.44c per contract
- Maker orders are 4x cheaper - strongly prefer GTC or POST_ONLY limit orders

## Strategy Guidelines
1. **Edge calculation**: Only trade when expected value > fees. At 50c with 7% taker fee, you need >53.5% edge to profit
2. **Maker orders**: Use limit orders with GTC or POST_ONLY to get the 1.75% maker rate
3. **Position sizing**: Don't concentrate more than 20% of equity in one position
4. **Settlement awareness**: Contracts settle to $1 or $0 at close - factor time remaining into trades
5. **Spread capture**: Consider placing limit orders inside the spread to capture the bid-ask

## Decision Process
Each step:
1. Call get_markets() to see current prices
2. Evaluate if any prices seem mispriced vs your probability estimates
3. Check your positions with get_positions() before trading
4. If you see opportunity, use limit orders (GTC) to get maker fees
5. Call done() when finished deciding for this step

Be conservative - the fees are high. Only trade when you see clear edge.
\end{verbatim}